\documentclass[journal]{IEEEtran}
\IEEEoverridecommandlockouts
\usepackage{lineno}
\usepackage{cite}
\usepackage{hyperref}
\usepackage{amsmath,amssymb,amsfonts}
\usepackage{amsmath}
\usepackage{amsthm}
\DeclareMathOperator*{\argmax}{argmax}
\usepackage{graphicx}
\usepackage{textcomp}
\usepackage{xcolor}
\usepackage{graphicx}
\usepackage{float}
\usepackage{subfigure}
\usepackage{amsmath}
\usepackage{amsfonts,amssymb}
\usepackage{mathrsfs}
\usepackage{mathtools}
\usepackage{algorithm}
\usepackage{algorithmicx}
\usepackage{algpseudocode}
\usepackage{bm}
\usepackage{multirow}
\usepackage{array}
\usepackage{amssymb}
\usepackage{amsmath}
\usepackage{cite}
\usepackage{url}
\usepackage{xcolor}
\usepackage{cite,graphicx,amsmath,amssymb}
\usepackage{subfigure}
\usepackage{fancyhdr}
\usepackage{mdwmath}
\usepackage{mdwtab}
\usepackage{caption}
\usepackage{amsthm}
\usepackage{setspace}
\usepackage{bm}
\usepackage{algorithm}
\usepackage{algpseudocode}
\usepackage{mathtools}
\usepackage{dsfont}
\usepackage{bbm}
\usepackage{cases}
\newtheorem{remark}{Remark}
\theoremstyle{definition}
\newtheorem{theorem}{Theorem}

\newtheorem{lemma}{Lemma}

\newtheorem{corollary}{Corollary}

\makeatletter
\newcommand{\biggg}{\bBigg@{3}}
\newcommand{\Biggg}{\bBigg@{3.5}}
\makeatother
\begin{document}
\title{Revealing the Impact of Beamforming in ISAC}
\author{Chongjun~Ouyang, Yuanwei~Liu, and Xingqi~Zhang
\thanks{C. Ouyang is with the School of Information and Communication Engineering, Beijing University of Posts and Telecommunications, Beijing, 100876, China (e-mail: DragonAim@bupt.edu.cn).}
\thanks{Y. Liu is with the School of Electronic Engineering and Computer Science, Queen Mary University of London, London, E1 4NS, U.K. (e-mail: yuanwei.liu@qmul.ac.uk).}
\thanks{X. Zhang is with the Department of Electrical and Computer Engineering, University of Alberta, Edmonton, T6G 2H5, Canada (e-mail: xingqi.zhang@ualberta.ca).}
}
\maketitle

\begin{abstract}
This letter proposes advanced beamforming design and analyzes its influence on the sensing and communications (S\&C) performance for a multiple-antenna integrated S\&C (ISAC) system with a single communication user and a single target. Novel closed-form beamformers are derived for three typical scenarios, including the sensing-centric design, communications-centric design, and Pareto optimal design. Regarding each scenario, the outage probability, ergodic communication rate (CR), and sensing rate (SR) are analyzed to derive the diversity orders and high signal-to-noise ratio slopes. Numerical results are provided to demonstrate that \romannumeral1) beamforming design can affect the high-SNR power offset and diversity order but does not influence the high-SNR slope; \romannumeral2) ISAC exhibits larger high-SNR slopes and a more extensive SR-CR region than conventional frequency-division S\&C (FDSAC) techniques.
\end{abstract}


\section{Introduction}
Integrated sensing and communications (ISAC) is a cutting-edge paradigm that enables seamless sharing of time-frequency-power-hardware resources between sensing and communications (S\&C) functionalities \cite{Ouyang2023_MCOM}. In contrast to the conventional frequency-division S\&C (FDSAC) techniques, where S\&C operations necessitate isolated frequency bands and dedicated hardware infrastructures, ISAC offers a more spectrum-, energy-, and hardware-efficient solution \cite{Zhang2022}. As a result, ISAC has garnered significant attention from both the academic community and industrial sectors alike \cite{Ouyang2023_MCOM,Zhang2022,Zhou2022,Wei2023}.

In recent times, considerable attention has been given to the application of multiple-antenna techniques in ISAC systems, as they offer significant beamforming gains that can benefit both S\&C functionalities; see \cite{Zhang2022,Zhou2022,Wei2023} and related references. However, the existing research primarily focuses on optimizing the beamforming design, and there is a notable lack of quantitative analysis concerning the fundamental impact of beamforming design on the overall S\&C performance. On the other hand, the assessment of information-theoretic limits in S\&C systems can be achieved through the S\&C mutual information (MI), which quantifies the amount of environmental or data information that can be recovered \cite{Ouyang2023_MCOM}. These performance metrics have been instrumental in establishing an upper limit on ISAC's performance capabilities \cite{Ouyang2023_MCOM}.

Despite their importance, there have been limited efforts to investigate the impact of beamforming design in ISAC systems from the perspective of MI, with the aim of enhancing S\&C performance. Although some studies have explored the influence of power allocation on S\&C MI, the specific effect of beamforming has not been fully addressed in these analyses \cite{Ouyang2023_WCL,Yuan2021}. Thus, the full extent of the relationship between beamforming and S\&C MI remains relatively unexplored.


To address the existing gap in the literature, this letter aims at investigating the beamforming design in ISAC systems from a MI perspective, with a particular focus on analyzing the impact of beamforming design on the S\&C performance. As an initial attempt, we consider a downlink multiple-antenna ISAC system with a single communication user (CU) and a single target.

The primary contributions of this letter are listed as follows: \romannumeral1) We propose a novel dual-functional S\&C (DFSAC) beamforming design tailored to three typical scenarios: the sensing-centric (S-C) design that maximizes the sensing rate (SR, the sensing MI each time-frequency unit), the communications-centric (C-C) design that maximizes the communication rate (CR, the communication MI each time-frequency unit), and the Pareto optimal design (characterizing the Pareto boundary of the SR-CR region). \romannumeral2) We derive optimal beamformers for each scenario in closed form, providing explicit expressions for the beamforming strategies to achieve the desired objectives. \romannumeral3) For each scenario, we analyze the outage probability (OP) of the CR and demonstrate that beamforming has no influence on the diversity order. \romannumeral4) For each scenario, we derive closed-form expressions for both the SR and ergodic CR (ECR), along with their approximations in the high signal-to-noise ratio (SNR) region, and establish that beamforming design does not affect the high-SNR slope but can significantly influence the high-SNR power offset. \romannumeral5) Through rigorous analysis, we demonstrate that ISAC provides superior degrees of freedom and a broader SR-CR region compared to the conventional FDSAC techniques.
\section{System Model}
We consider an ISAC system as sketched in {\figurename} {\ref{figure1}}, where a DFSAC base station (BS) is serving a single-antenna CU, while simultaneously sensing a single target in its surrounding environment. The BS is equipped with $M$ transmit antennas and $N$ receive antennas. Let ${\mathbf{X}}=[{\mathbf{x}}_1\ldots{\mathbf{x}}_L]\in{\mathbbmss{C}}^{M\times L}$ be a DFSAC signal matrix, with $L$ being the length of the communication frame/sensing pulse. From a communication perspective, ${\mathbf{x}}_l\in{\mathbbmss{C}}^{M\times 1}$ for $l\in{\mathcal{L}}=\{1,\ldots,L\}$ denotes the $l$th data symbol vector. For sensing, ${\mathbf{x}}_l$ represents the sensing snapshot transmitted at the $l$th time slot. In the considered ISAC system, we could design the signal matrix as ${\mathbf{X}}=\sqrt{p}{\mathbf{w}}\mathbf{s}^{\mathsf{H}}$, where $\mathbf{w}\in{\mathbbmss{C}}^{M\times1}$ is the normalized beamforming vector, $p$ is the power budget, and $\mathbf{s}\in{\mathbbmss{C}}^{L\times1}$ denotes the unit-power data streams intended for the CU with $L^{-1}\lVert{\mathbf{s}}\rVert^2=1$.
\subsection{Communication Model}
\begin{figure}[!t]
 \centering
\setlength{\abovecaptionskip}{0pt}
\includegraphics[height=0.13\textwidth]{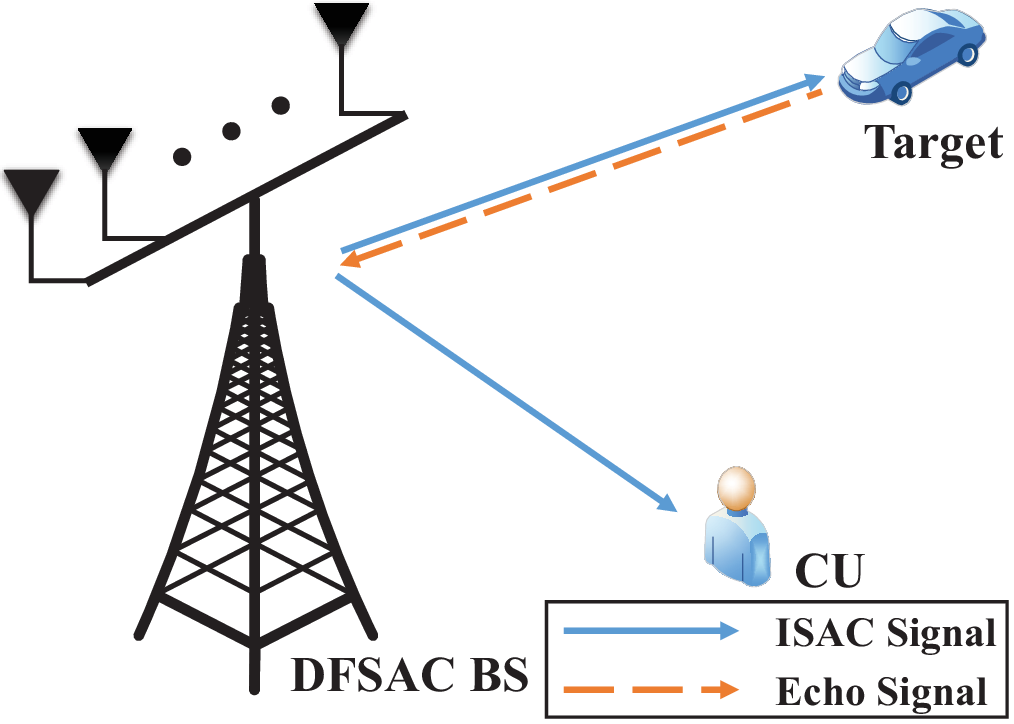}
\caption{Illustration of an ISAC system.}
    \label{figure1}
    \vspace{-20pt}
\end{figure}

The observation at the CU can be written as follows:
{\setlength\abovedisplayskip{2pt}
\setlength\belowdisplayskip{2pt}
\begin{align}
{{\mathbf{y}}_{\rm{c}}^{\mathsf{H}}}={\mathbf{h}}_{\rm{c}}^{\mathsf{H}}{\mathbf{X}}+{\mathbf{n}}_{\rm{c}}^{\mathsf{H}}
=\sqrt{p}{\mathbf{h}}_{\rm{c}}^{\mathsf{H}}{\mathbf{w}}\mathbf{s}^{\mathsf{H}}
+{{\mathbf{n}}_{\rm{c}}^{\mathsf{H}}},
\end{align}
}where ${\mathbf{n}}_{\rm{c}}\sim{\mathcal{CN}}({\mathbf{0}},{\mathbf{I}})$ is the additive white Gaussian noise (AWGN) vector, and ${\mathbf{h}}_{\rm{c}}\in{\mathbbmss C}^{M\times1}$ represents the communication channel, which is assumed to be known to the BS. We consider that the communication link shown in {\figurename} {\ref{figure1}} suffers Rayleigh fading, which yields ${\mathbf{h}}_{\rm{c}}\sim{\mathcal{CN}}({\mathbf{0}},\alpha_{\rm{c}}{\mathbf{I}})$ with $\alpha_{\rm{c}}$ reflecting the influence of the large-scale path loss. Accordingly, the received SNR at the CU is given by $\gamma_{\rm{c}}=p\lvert{\mathbf{w}}^{\mathsf{H}}{\mathbf{h}}_{\rm{c}}\rvert^2$. It follows that the CR satisfies ${\mathcal{R}}_{\rm{c}}=\log_2(1+\gamma_{\rm{c}})$.
\subsection{Sensing Model}
By transmitting $\mathbf{X}$ to sense the target, the BS observes the following reflected echo signal matrix at its receiver:
{\setlength\abovedisplayskip{2pt}
\setlength\belowdisplayskip{2pt}
\begin{align}\label{reflected_echo_signal_matrix}
{\mathbf{Y}}_{\rm{s}}={\mathbf{G}}{\mathbf{X}}+{\mathbf{N}}_{\rm{s}}=
\sqrt{p}{\mathbf{G}}{\mathbf{w}}\mathbf{s}^{\mathsf{H}}+{\mathbf{N}}_{\rm{s}},
\end{align}
}where ${\mathbf{N}}_{\rm{s}}\in{\mathbbmss{C}}^{N\times L}$ is the AWGN matrix with each entry having zero mean and unit variance, and $\mathbf{G}\in{\mathbbmss{C}}^{N\times M}$ represents the target response matrix. Specifically, the target response matrix can be modeled by \cite{Ouyang2022_CL,Ouyang2023_MCOM,Ouyang2023_WCL}
{\setlength\abovedisplayskip{2pt}
\setlength\belowdisplayskip{2pt}
\begin{align}\label{Target_Response_Matrix_Formula}
{\mathbf{G}}=\beta{\mathbf{a}}\left(\theta\right){\mathbf{b}}^{\mathsf{H}}\left(\theta\right),
\end{align}
}where $\beta$ is the complex amplitude of this single target, ${\mathbf{a}}\left(\theta\right)\in{\mathbbmss{C}}^{N\times1}$ and ${\mathbf{b}}\left(\theta\right)\in{\mathbbmss{C}}^{M\times1}$ are the associated receive and transmit array steering vectors, respectively, and $\theta$ is the angle of the target. Following the Swerling target model, we model the reflection coefficient of the target as a complex Gaussian variate, i.e., $\beta\sim{\mathcal{CN}}(0,\alpha_{\rm{s}})$, where $\alpha_{\rm{s}}$ represents the average strength.

We consider that the angle of the target is perfectly tracked and focus on the estimation of the reflection coefficient $\beta$. Under this circumstance, the sensing task aims at extracting the environmental information contained in $\beta$ from ${\mathbf{Y}}_{\rm{s}}$ by knowing $\mathbf{X}$. The information-theoretic limits on this sensing task is thus characterized by the sensing MI, which is defined as the MI between ${\mathbf{Y}}_{\rm{s}}$ and $\beta$ or $\mathbf{G}$ conditioned on the DFSAC signal ${\mathbf{X}}$ \cite{Ouyang2022_CL,Ouyang2023_MCOM,Ouyang2023_WCL}. On this basis, we adopt the SR as the performance metric of sensing, which is defined as the sensing MI per unit time \cite{Ouyang2022_CL,Ouyang2023_MCOM,Ouyang2023_WCL}. Assuming that each DFSAC symbol lasts 1 unit time, we write the SR as follows:
{\setlength\abovedisplayskip{2pt}
\setlength\belowdisplayskip{2pt}
\begin{align}
{\mathcal{R}}_{\rm{s}}=L^{-1}I\left({\mathbf{Y}}_{\rm{s}};\beta|\mathbf{X}\right)=L^{-1}I\left({\mathbf{Y}}_{\rm{s}};\mathbf{G}|\mathbf{X}\right),
\end{align}
}where $I\left(X;Y|Z\right)$ denotes the MI between $X$ and $Y$ conditioned on $Z$. For a given $\mathbf{w}$, we calculate ${\mathcal{R}}_{\rm{s}}$ as follows.
\vspace{-5pt}
\begin{lemma}\label{SR_Def_Lemma}
The SR can be calculated as follows:
{\setlength\abovedisplayskip{2pt}
\setlength\belowdisplayskip{2pt}
\begin{align}\label{SR_Basic_Exp}
{\mathcal{R}}_{\rm{s}}=L^{-1}\log_2(1+pNL\alpha_{\rm{s}}\lvert{\mathbf{w}}^{\mathsf{H}}{\mathbf{h}}_{\rm{s}}\rvert^2),
\end{align}
}where ${\mathbf{h}}_{\rm{s}}\triangleq{\mathbf{b}}\left(\theta\right)$ can be treated as the sensing channel.
\end{lemma}
\vspace{-5pt}
\begin{IEEEproof}
Please refer to Appendix \ref{Proof_SR_Def_Lemma} for more details.
\end{IEEEproof}
Given the above ISAC framework, we intend to analyze its S\&C performance by investigating the CR $\mathcal{R}_{\rm{c}}$ and SR $\mathcal{R}_{\rm{s}}$. Note that both $\mathcal{R}_{\rm{c}}$ and $\mathcal{R}_{\rm{s}}$ are influenced by the beamforming vector $\mathbf{w}$. However, finding an optimal $\mathbf{w}$ that maximizes both $\mathcal{R}_{\rm{s}}$ and $\mathcal{R}_{\rm{c}}$ concurrently poses a challenging task. Motivated by this challenge, we explore three distinct scenarios to gain further insights into the system, i.e., the S-C design, the C-C design, and the Pareto optimal design.
\section{Performance of ISAC}
\subsection{Communications-Centric Design}
Under the C-C design, the beamforming vector $\mathbf{w}$ is set to maximize ${\mathcal{R}}_{\rm{c}}$, which satisfies
{\setlength\abovedisplayskip{2pt}
\setlength\belowdisplayskip{2pt}
\begin{align}\label{C_C_Beamforming_Vector}
\argmax\nolimits_{{\mathbf{w}}}{\mathcal{R}}_{\rm{c}}
=\argmax\nolimits_{{\mathbf{w}}}\lvert{\mathbf{w}}^{\mathsf{H}}{\mathbf{h}}_{\rm{c}}\rvert
={\lVert{\mathbf{h}}_{\rm{c}}\rVert}^{-1}{\mathbf{h}}_{\rm{c}}\triangleq{\mathbf{w}}_{\rm{c}}.
\end{align}
}\subsubsection{Performance of Communications}
Given ${\mathbf{w}}={\mathbf{w}}_{\rm{c}}$, the CR can be written as ${\overline{\mathcal{R}}}_{{\rm{c}}}^{\rm{c}}=\log_2\left(1+p\lVert{\mathbf{h}}_{\rm{c}}\rVert^2\right)$. We next exploit the OP and the ECR to evaluate the performance of communications. Theorem \ref{Ergodic_Communication_Rate_C_C_Theorem} provides an exact expression for the ECR ${{\mathcal{R}}}_{{\rm{c}}}^{\rm{c}}={\mathbbmss{E}}\{{\overline{\mathcal{R}}}_{{\rm{c}}}^{\rm{c}}\}$ and its high-SNR approximation.
\vspace{-5pt}
\begin{theorem}\label{Ergodic_Communication_Rate_C_C_Theorem}
In the C-C design, the ECR is given by
{\setlength\abovedisplayskip{2pt}
\setlength\belowdisplayskip{2pt}
\begin{align}
{{\mathcal{R}}}_{{\rm{c}}}^{\rm{c}}&=\sum\nolimits_{\mu=0}^{M-1}\frac{(-1/(p\alpha_{\rm{c}}))^{M-1-\mu}}{(M-1-\mu)!\ln{2}}
\left(-{\rm{e}}^{\frac{1}{p\alpha_{\rm{c}}}}{\rm{Ei}}\left(\frac{-1}{p\alpha_{\rm{c}}}\right)\right.\nonumber\\
&+\left.\sum\nolimits_{i=1}^{M-1-\mu}(i-1)!\left(-1/(p\alpha_{\rm{c}})\right)^{-i}\right),\label{Ergodic_Communication_Rate_S_C_Basic}
\end{align}
}where ${\rm{Ei}}(x)=-\int_{-x}^{\infty}{\rm{e}}^{-t}t^{-1}{\rm{d}}t$ is the exponential integral function \cite[Eq. (8.211.1)]{Ryzhik2007}. As $p\rightarrow\infty$, the ECR satisfies
{\setlength\abovedisplayskip{2pt}
\setlength\belowdisplayskip{2pt}
\begin{align}\label{Ergodic_Communication_Rate_C_C_Asymptotic}
{{\mathcal{R}}}_{{\rm{c}}}^{\rm{c}}
\approx \log_2{p}+\log_2{\alpha_{\rm{c}}}+\psi\left(M\right)/\ln{2},
\end{align}
}where $\psi\left(x\right)=\frac{{\rm d}}{{\rm d}x}\ln{\Gamma\left(x\right)}$ is the Digamma function \cite[Eq. (6.461)]{Ryzhik2007} and $\Gamma\left(x\right)=\int_{0}^{\infty}t^{x-1}{\rm e}^{-t}{\rm d}t$ is the gamma function \cite[Eq. (6.1.1)]{Ryzhik2007}. When $x\in{\mathbbmss{Z}}^{+}$, $\psi(x)=\psi(1)+\sum_{i=1}^{x-1}\frac{1}{i}$.
\end{theorem}
\vspace{-5pt}
\begin{IEEEproof}
Please refer to Appendix \ref{Proof_Ergodic_Communication_Rate_C_C_Theorem} for more details.
\end{IEEEproof}
\vspace{-5pt}
\begin{remark}
The results in \eqref{Ergodic_Communication_Rate_C_C_Asymptotic} suggest that the high-SNR slope and the high-SNR power offset of ${{\mathcal{R}}}_{{\rm{c}}}^{\rm{c}}$ are given by ${{\mathcal{S}}}_{{\rm{c}}}^{\rm{c}}=1$ and ${{\mathcal{L}}}_{{\rm{c}}}^{\rm{c}}=-\log_2{\alpha_{\rm{c}}}-\frac{\psi\left(M\right)}{\ln{2}}$, respectively.
\end{remark}
\vspace{-5pt}
Turn to the OP ${{\mathcal{P}}}_{{\rm{c}}}^{\rm{c}}=\Pr({\overline{\mathcal{R}}}_{{\rm{c}}}^{\rm{c}}<\mathcal{R}_0)$, where $\mathcal{R}_0$ denotes the target CR. The following theorem provides an exact expression for the OP as well as its high-SNR approximation.
\vspace{-5pt}
\begin{theorem}\label{Outage_Probability_UT_C_C_Theorem}
In the C-C design, the OP is given by ${{\mathcal{P}}}_{{\rm{c}}}^{\rm{c}}=\frac{1}{\Gamma(M)}\gamma(M,\frac{2^{\mathcal{R}_0}-1}{p\alpha_{\rm{c}}})$, where $\gamma\left(s,x\right)=\int_{0}^{x}t^{s-1}{\rm e}^{-t}{\rm d}t$ is the lower incomplete gamma function \cite[Eq. (8.350.1)]{Ryzhik2007}. As $p\rightarrow\infty$, the ECR satisfies ${{\mathcal{P}}}_{{\rm{c}}}^{\rm{c}}\approx\frac{(2^{\mathcal{R}_0}-1)^M}{p^M\alpha_{\rm{c}}^MM!}$.
\end{theorem}
\vspace{-5pt}
\begin{IEEEproof}
Please refer to Appendix \ref{Proof_Ergodic_Communication_Rate_C_C_Theorem} for more details.
\end{IEEEproof}
\vspace{-5pt}
\begin{remark}
Theorem \ref{Outage_Probability_UT_C_C_Theorem} suggests that a diversity order of ${{\mathcal{D}}}_{{\rm{c}}}^{\rm{c}}=M$ is achievable for the OP under the C-C design.
\end{remark}
\vspace{-5pt}
\subsubsection{Performance of Sensing}
For ${\mathbf w}={\mathbf{w}}_{\rm{c}}$, the SR reads
{\setlength\abovedisplayskip{2pt}
\setlength\belowdisplayskip{2pt}
\begin{align}
{\overline{\mathcal R}}_{\rm{s}}^{\rm{c}}=L^{-1}\log_2(1+pNL\alpha_{\rm{s}}\lvert{\mathbf{h}}_{\rm{c}}^{\mathsf{H}}{\mathbf{h}}_{\rm{s}}\rvert^2
{\lVert{\mathbf{h}}_{\rm{c}}\rVert}^{-2}).
\end{align}
}Noticing the statistics of $\mathbf{h}_{\rm{c}}$, we further define the average SR as ${{\mathcal R}}_{\rm{s}}^{\rm{c}}={\mathbbmss E}\{{\overline{\mathcal R}}_{\rm{s}}^{\rm{c}}\}$, which can be calculated numerically. Besides, the following theorem is found.
\vspace{-5pt}
\begin{theorem}\label{Average_Sensing_Rate_C_C_Theorem}
As $p\rightarrow\infty$, ${{\mathcal R}}_{\rm{s}}^{\rm{c}}$ can be approximated as follows:
{\setlength\abovedisplayskip{2pt}
\setlength\belowdisplayskip{2pt}
\begin{align}\label{Average_Sensing_Rate_C_C_Asymptotic}
{{\mathcal{R}}}_{{\rm{s}}}^{\rm{c}}
\approx \frac{1}{L}\left(\log_2{p}+\log_2\left(LN\alpha_{\rm{s}}\lVert{\mathbf{h}}_{\rm{s}}\rVert^2\right)
\!-\!\sum_{i=1}^{M-1}\frac{1}{i\ln{2}}\right).
\end{align}
}\end{theorem}
\vspace{-5pt}
\begin{IEEEproof}
Please refer to Appendix \ref{Proof_Average_Sensing_Rate_C_C_Theorem} for more details.
\end{IEEEproof}
\vspace{-5pt}
\begin{remark}
The results in \eqref{Average_Sensing_Rate_C_C_Asymptotic} suggest that the high-SNR slope and the high-SNR power offset of the SR achieved by the C-C design are given by ${{\mathcal{S}}}_{{\rm{s}}}^{\rm{c}}=\frac{1}{L}$ and ${{\mathcal{L}}}_{{\rm{s}}}^{\rm{c}}=-\log_2\left(LN\alpha_{\rm{s}}\lVert{\mathbf{h}}_{\rm{s}}\rVert^2\right)
+\sum_{i=1}^{M-1}\frac{1}{i\ln{2}}$, respectively.
\end{remark}
\vspace{-5pt}
\subsection{Sensing-Centric Design}
Under the S-C design, the beamforming vector $\mathbf{w}$ is set to maximize ${\mathcal{R}}_{\rm{s}}$, which satisfies
{\setlength\abovedisplayskip{2pt}
\setlength\belowdisplayskip{2pt}
\begin{align}\label{S_C_Beamforming_Vector}
\argmax\nolimits_{{\mathbf{w}}}{\mathcal{R}}_{\rm{s}}
=\argmax\nolimits_{{\mathbf{w}}}\lvert{\mathbf{w}}^{\mathsf{H}}{\mathbf{h}}_{\rm{s}}\rvert
={\lVert{\mathbf{h}}_{\rm{s}}\rVert}^{-1}{\mathbf{h}}_{\rm{s}}\triangleq{\mathbf{w}}_{\rm{s}}.
\end{align}
}\subsubsection{Performance of Sensing}
The following theorem provides an exact expression for the SR achieved by the S-C design as well as its high-SNR approximation.
\vspace{-5pt}
\begin{theorem}\label{Sensing_Rate_S_C_Theorem}
In the S-C design, the achieved SR is given by
{\setlength\abovedisplayskip{2pt}
\setlength\belowdisplayskip{2pt}
\begin{align}
\mathcal{R}_{\rm{s}}^{\rm{s}}=L^{-1}\log_2(1+pNL\alpha_{\rm{s}}\lVert{\mathbf{h}}_{\rm{s}}\rVert^2).
\end{align}
}As $p\rightarrow\infty$, the SR can be approximated as follows:
{\setlength\abovedisplayskip{2pt}
\setlength\belowdisplayskip{2pt}
\begin{align}\label{Sensing_Rate_S_C_Asymptotic}
\mathcal{R}_{\rm{s}}^{\rm{s}}
\approx L^{-1}\left(\log_2{p}+\log_2\left(NL\alpha_{\rm{s}}\lVert{\mathbf{h}}_{\rm{s}}\rVert^2\right)\right).
\end{align}
}\end{theorem}
\vspace{-5pt}
\begin{IEEEproof}
Similar to the proof of Theorem \ref{Average_Sensing_Rate_C_C_Theorem}.
\end{IEEEproof}
\vspace{-5pt}
\begin{remark}
The results in \eqref{Sensing_Rate_S_C_Asymptotic} suggest that the high-SNR slope and the high-SNR power offset of $\mathcal{R}_{\rm{s}}^{\rm{s}}$ are given by ${{\mathcal{S}}}_{{\rm{s}}}^{\rm{s}}=\frac{1}{L}$ and ${{\mathcal{L}}}_{{\rm{s}}}^{\rm{s}}=-\log_2\left(LN\alpha_{\rm{s}}\lVert{\mathbf{h}}_{\rm{s}}\rVert^2\right)$, respectively.
\end{remark}
\vspace{-5pt}
\vspace{-5pt}
\begin{remark}\label{BF_Influence_Sensing_1}
The fact of ${{\mathcal{S}}}_{{\rm{s}}}^{\rm{s}}={{\mathcal{S}}}_{{\rm{s}}}^{\rm{c}}=L^{-1}$ means that the beamforming design does not influence the high-SNR slope of the SR. By contrast, the fact of ${{\mathcal{L}}}_{{\rm{s}}}^{\rm{c}}-{{\mathcal{L}}}_{{\rm{s}}}^{\rm{s}}=\sum_{i=1}^{M-1}\frac{1}{i\ln{2}}$ suggests that the beamforming design affects the SR via shaping its high-SNR power offset, and the sensing performance gap between the S-C design and the C-C design is more highlighted when the DFSAC BS has more transmit antennas.
\end{remark}
\vspace{-5pt}
\subsubsection{Performance of Communications}
The CR achieved by the S-C design can be written as ${\overline{\mathcal{R}}}_{{\rm{c}}}^{\rm{s}}=\log_2\left(1+p\lvert{\mathbf{w}}_{\rm{s}}^{\mathsf{H}}{\mathbf{h}}_{\rm{c}}\rvert^2\right)$. The following theorem provides an exact expression for the ECR ${{\mathcal{R}}}_{{\rm{c}}}^{\rm{s}}={\mathbbmss{E}}\{{\overline{\mathcal{R}}}_{{\rm{c}}}^{\rm{s}}\}$ as well as its high-SNR approximation.
\vspace{-5pt}
\begin{theorem}\label{Ergodic_Communication_Rate_S_C_Theorem}
In the S-C design, the ECR is given by
{\setlength\abovedisplayskip{2pt}
\setlength\belowdisplayskip{2pt}
\begin{align}
{{\mathcal{R}}}_{{\rm{c}}}^{\rm{s}}=-{\rm{e}}^{1/p/\alpha_{\rm{c}}}{\rm{Ei}}(-1/p/\alpha_{\rm{c}})/\ln{2}.\label{Ergodic_Communication_Rate_S_C_Basic}
\end{align}
} As $p\rightarrow\infty$, the ECR satisfies
{\setlength\abovedisplayskip{2pt}
\setlength\belowdisplayskip{2pt}
\begin{align}\label{Ergodic_Communication_Rate_S_C_Asymptotic}
{{\mathcal{R}}}_{{\rm{c}}}^{\rm{s}}
\approx \log_2{p}+\log_2{\alpha_{\rm{c}}}+\psi\left(1\right)/\ln{2}.
\end{align}
}\end{theorem}
\vspace{-5pt}
\begin{IEEEproof}
Please refer to Appendix \ref{Proof_Ergodic_Communication_Rate_C_C_Theorem} for more details.
\end{IEEEproof}
\vspace{-5pt}
\begin{remark}
The results in \eqref{Ergodic_Communication_Rate_S_C_Asymptotic} suggest that the high-SNR slope and the high-SNR power offset of ${{\mathcal{R}}}_{{\rm{c}}}^{\rm{s}}$ are given by ${{\mathcal{S}}}_{{\rm{c}}}^{\rm{s}}=1$ and ${{\mathcal{L}}}_{{\rm{c}}}^{\rm{s}}=-\log_2{\alpha_{\rm{c}}}-\frac{\psi\left(1\right)}{\ln{2}}$, respectively.
\end{remark}
\vspace{-5pt}
Turn to the OP ${{\mathcal{P}}}_{{\rm{c}}}^{\rm{s}}=\Pr({\overline{\mathcal{R}}}_{{\rm{c}}}^{\rm{s}}<\mathcal{R}_0)$. The following theorem provides an exact expression for the OP.
\vspace{-5pt}
\begin{theorem}\label{Outage_Probability_UT_S_C_Theorem}
In the S-C design, the OP is given by ${{\mathcal{P}}}_{{\rm{c}}}^{\rm{s}}=1-{\rm{e}}^{-\frac{2^{\mathcal{R}_0}-1}{p\alpha_{\rm{c}}}}$. As $p\rightarrow\infty$, the ECR satisfies ${{\mathcal{P}}}_{{\rm{c}}}^{\rm{s}}\approx\frac{2^{\mathcal{R}_0}-1}{p\alpha_{\rm{c}}}$.
\end{theorem}
\vspace{-5pt}
\begin{IEEEproof}
Please refer to Appendix \ref{Proof_Ergodic_Communication_Rate_C_C_Theorem} for more details.
\end{IEEEproof}
\vspace{-5pt}
\begin{remark}
Theorem \ref{Outage_Probability_UT_S_C_Theorem} suggests that a diversity order of ${{\mathcal{D}}}_{{\rm{c}}}^{\rm{s}}=1$ is achievable for the OP under the S-C design.
\end{remark}
\vspace{-5pt}
\vspace{-5pt}
\begin{remark}\label{BF_Influence_Communications_1}
The fact of ${{\mathcal{S}}}_{{\rm{c}}}^{\rm{s}}={{\mathcal{S}}}_{{\rm{c}}}^{\rm{c}}=1$ means that the beamforming design does not influence the high-SNR slope of the CR. By contrast, the fact of ${{\mathcal{D}}}_{{\rm{c}}}^{\rm{s}}={{\mathcal{D}}}_{{\rm{c}}}^{\rm{c}}/M$ and ${{\mathcal{L}}}_{{\rm{c}}}^{\rm{s}}-{{\mathcal{L}}}_{{\rm{c}}}^{\rm{c}}=\frac{\psi(M)-\psi(1)}{\ln{2}}=\sum_{i=1}^{M-1}\frac{1}{i\ln{2}}$ suggests that the beamforming design affects the CR via shaping its high-SNR power offset and diversity order, and the communication performance gap between the S-C design and the C-C design is more highlighted when the DFSAC BS has more transmit antennas.
\end{remark}
\vspace{-5pt}
\subsection{Pareto Optimal Design}
\begin{figure}[!t]
 \centering
\setlength{\abovecaptionskip}{0pt}
\includegraphics[height=0.13\textwidth]{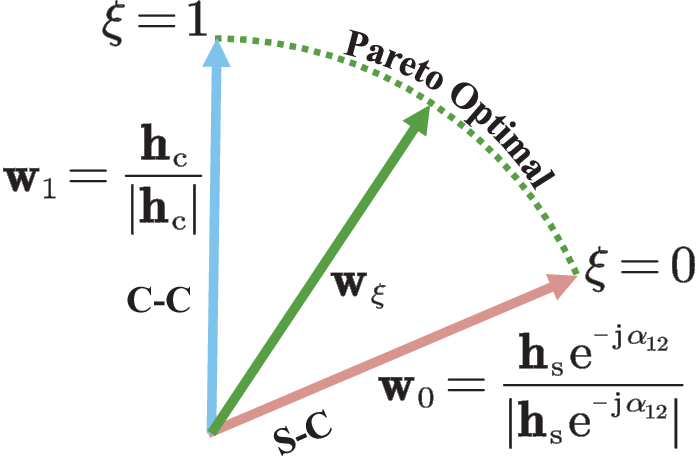}
\caption{Pareto optimal beamformer for problem \eqref{Problem_CR_SR_Tradeoff}.}
    \label{Trade-Off_Legend}
    \vspace{-20pt}
\end{figure}
In addition to maximizing the CR or SR, the beamforming vector $\mathbf{w}$ can be designed to satisfy different qualities of services, which results in a communication-sensing performance tradeoff. This tradeoff can be evaluated by the Pareto boundary of the SR-CR region. Particularly, any rate-tuple on the Pareto boundary can be obtained via the rate-profile based method, i.e., solving the problem as follows \cite{Ouyang2023_WCL}:
{\setlength\abovedisplayskip{2pt}
\setlength\belowdisplayskip{2pt}
\begin{equation}\label{Problem_CR_SR_Tradeoff}
\max\nolimits_{{\mathbf{w}},\mathcal{R}}{\mathcal{R}},~{\rm{s.t.}}~{\mathcal{R}}_{\rm{s}}\geq \alpha{\mathcal{R}},{\mathcal{R}}_{\rm{c}}\geq (1-\alpha){\mathcal{R}},\lVert{\mathbf{w}}\rVert^2=1,
\end{equation}
}where $\alpha\in[0,1]$ is a rate-profile parameter. The entire Pareto boundary is obtained by solving the above problem with $\alpha$ varying from $0$ to $1$. Despite of its non-convexity, problem \eqref{Problem_CR_SR_Tradeoff} can be optimally solved in a closed form as follows.
\vspace{-5pt}
\begin{theorem}\label{Optimal_Solution_Problem_CR_SR_Tradeoff}
Given $\alpha$, we denote $\beta_1=(1-\alpha)\ln{2}$, $\beta_2={L\alpha}\ln{2}$, ${\mathbf{h}}_1=\sqrt{p}{\mathbf{h}}_{\rm{c}}$, ${\mathbf{h}}_2=\sqrt{pNL\alpha_{\rm{s}}}{\mathbf{h}}_{\rm{s}}$, $\alpha_{11}=p\lVert{\mathbf{h}}_{\rm{c}}\rVert^2$, $\alpha_{22}=pNL\alpha_{\rm{s}}\lVert{\mathbf{h}}_{\rm{s}}\rVert^2$, $\alpha_{12}=p\sqrt{NL\alpha_{\rm{s}}}{\mathbf{h}}_{\rm{c}}^{\mathsf{H}}{\mathbf{h}}_{\rm{s}}$, $\delta=\sqrt{\frac{{\rm{e}}^{\beta_2{\mathcal{R}}}-1}{{\rm{e}}^{\beta_1{\mathcal{R}}}-1}}$, $\varrho_1=\alpha_{22}-\delta|\alpha_{12}|$,
$\varrho_2=\alpha_{11}-\delta^{-1}|\alpha_{12}|$, $\chi=\varrho_1\beta_1{\rm{e}}^{\beta_1{\mathcal{R}}}+\varrho_2\beta_2{\rm{e}}^{\beta_2{\mathcal{R}}}$, $\mu_1=\frac{\varrho_1}{\chi}$, $\mu_2=\frac{\varrho_2}{\chi}$. Let ${\mathcal{R}}^{\star}$ denote the solution of $\mathcal{R}$ to the equation $\alpha_{11}\alpha_{22}-|\alpha_{12}|^2=(\alpha_{11}-\delta^{-1}|\alpha_{12}|)({\rm{e}}^{\beta_2{\mathcal{R}}}-1)
+(\alpha_{22}-\delta|\alpha_{12}|)({\rm{e}}^{\beta_1{\mathcal{R}}}-1)$. Then, the optimal beamforming vector in problem \eqref{Problem_CR_SR_Tradeoff} is given by
{\setlength\abovedisplayskip{2pt}
\setlength\belowdisplayskip{2pt}
\begin{equation}
{\mathbf{w}}_\alpha^{\star}\!=\!\left\{
\begin{array}{ll}
\!\!\!{\mathbf{w}}_{\rm{c}}             & {\mu_2      =      0,\mu_1>0}\\
\!\!\!\frac{\tau\mu_1{\mathbf{h}}_{1}}{({{\rm{e}}^{\beta_1{\mathcal{R}}}-1})^{-0.5}}\!+\!
\frac{\tau\mu_2{\mathbf{h}}_{2}{\rm{e}}^{-{\rm{j}}\angle\alpha_{12}}}{({{\rm{e}}^{\beta_2{\mathcal{R}}}-1})^{-0.5}}           & {\mu_1>0,\mu_2>0}\\
\!\!\!{\mathbf{w}}_{\rm{s}}             & {\mu_1      =      0,\mu_2>0}
\end{array} \right.\!\!\!.
\end{equation}
}where ${\mathcal{R}}$ in $\{\delta,\chi\}$ satisfies ${\mathcal{R}}={\mathcal{R}}^{\star}$ and $\tau$ is for normalization.
\end{theorem}
\vspace{-5pt}
\begin{IEEEproof}
Please refer to Appendix \ref{Proof_Optimal_Solution_Problem_CR_SR_Tradeoff} for more details.
\end{IEEEproof}
\vspace{-5pt}
\begin{corollary}\label{Simplified_Solution_Problem_CR_SR_Tradeoff}
The whole Pareto boundary of the rate region can be achieved by the beamforming vector as follows:
{\setlength\abovedisplayskip{2pt}
\setlength\belowdisplayskip{2pt}
\begin{equation}
{\mathbf{w}}_{\xi}=\frac{\xi{\mathbf{h}}_{\rm{c}}+({1-\xi}){\mathbf{h}}_{\rm{s}}{\rm{e}}^{-{\rm{j}}\angle\alpha_{12}}}
{\lvert\xi{\mathbf{h}}_{\rm{c}}+({1-\xi}){\mathbf{h}}_{\rm{s}}{\rm{e}}^{-{\rm{j}}\angle\alpha_{12}}\rvert},
\end{equation}
}where the weighting factor $\xi$ varies between $[0,1]$.
\end{corollary}
\vspace{-5pt}
\begin{IEEEproof}
Please refer to Appendix \ref{Proof_Simplified_Solution_Problem_CR_SR_Tradeoff} for more details.
\end{IEEEproof}
\vspace{-5pt}
\begin{remark}\label{Subspace_Trade_Off}
Note that ${\mathbf{w}}_{\xi}$ can represent any arbitrary linear combination of ${\mathbf{h}}_{\rm{c}}$ and ${\mathbf{h}}_{\rm{s}}{\rm{e}}^{-{\rm{j}}\angle\alpha_{12}}$ with non-negative real coefficients. The results in Corollary \ref{Simplified_Solution_Problem_CR_SR_Tradeoff} suggest that the Pareto optimal beamforming vector lies in the plane spanned by ${\mathbf{h}}_{\rm{c}}$ and ${\mathbf{h}}_{\rm{s}}{\rm{e}}^{-{\rm{j}}\angle\alpha_{12}}$, as depicted in {\figurename} {\ref{Trade-Off_Legend}}.
\end{remark}
\vspace{-5pt}
Given $\alpha$, let ${{\overline{\mathcal{R}}}}_{\rm{c}}^{\alpha}$, ${{\mathcal{R}}}_{\rm{c}}^{\alpha}$, and ${{\mathcal{R}}}_{\rm{s}}^{\alpha}$ denote the instantaneous CR, ECR, and average SR achieved by ${\mathbf{w}}_\alpha^{\star}$, respectively. It follows that ${{\mathcal{R}}}_{\rm{c}}^{\alpha}\in\left[{{\mathcal{R}}}_{\rm{c}}^{\rm{s}},{{\mathcal{R}}}_{\rm{c}}^{\rm{c}}\right]$ and ${{\mathcal{R}}}_{\rm{s}}^{\alpha}\in\left[{{\mathcal{R}}}_{\rm{s}}^{\rm{c}},{{\mathcal{R}}}_{\rm{s}}^{\rm{s}}\right]$ with ${{\mathcal{R}}}_{\rm{s}}^{1}={{\mathcal{R}}}_{\rm{s}}^{\rm{s}}$ and ${{\mathcal{R}}}_{\rm{c}}^{0}={{\mathcal{R}}}_{\rm{c}}^{\rm{c}}$. By the Sandwich theorem, we find Corollary \ref{CR_SR_Pareto_Boundary_Asymptotic}.
\vspace{-5pt}
\begin{corollary}\label{CR_SR_Pareto_Boundary_Asymptotic}
For a sufficiently larger SNR, i.e., $p\rightarrow\infty$, ${{\mathcal{R}}}_{\rm{s}}^{\alpha}\approx{\mathcal{S}}_{\rm{s}}^{\alpha}(\log_2{p}-{\mathcal{L}}_{\rm{s}}^{\alpha})$, ${{\mathcal{R}}}_{\rm{c}}^{\alpha}\approx{\mathcal{S}}_{\rm{c}}^{\alpha}(\log_2{p}-{\mathcal{L}}_{\rm{c}}^{\alpha})$, and ${\Pr}({{\overline{\mathcal{R}}}}_{\rm{c}}^{\alpha}<{\mathcal{R}}_0)\simeq{\mathcal{O}}(p^{-{{\mathcal{D}}}_{\rm{c}}^{\alpha}})$, where ${\mathcal{S}}_{\rm{s}}^{\alpha}=\frac{1}{L}$, ${\mathcal{L}}_{\rm{s}}^{\alpha}\in[{{\mathcal{L}}}_{{\rm{s}}}^{\rm{s}},{{\mathcal{L}}}_{{\rm{s}}}^{\rm{c}}]$, ${\mathcal{S}}_{\rm{c}}^{\alpha}=1$, ${\mathcal{L}}_{\rm{c}}^{\alpha}\in[{{\mathcal{L}}}_{{\rm{c}}}^{\rm{c}},{{\mathcal{L}}}_{{\rm{c}}}^{\rm{s}}]$, and ${\mathcal{D}}_{\rm{c}}^{\alpha}\in[{{\mathcal{D}}}_{{\rm{c}}}^{\rm{s}},{{\mathcal{D}}}_{{\rm{c}}}^{\rm{c}}]$.
\end{corollary}
\vspace{-5pt}
\vspace{-5pt}
\begin{remark}
The arguments in Remark \ref{BF_Influence_Sensing_1}, Remark \ref{BF_Influence_Communications_1}, and Corollary \ref{CR_SR_Pareto_Boundary_Asymptotic} collectively suggest that the beamforming design influences the CR and SR via shaping the high-SNR power offset and diversity order rather than the high-SNR slope.
\end{remark}
\vspace{-5pt}
\vspace{-5pt}
\begin{remark}
Denote ${\mathcal{R}}^{\rm{s}}$ and ${\mathcal{R}}^{\rm{c}}$ as the achievable SR and CR, respectively. Then, the rate region achieved by ISAC reads
{\setlength\abovedisplayskip{2pt}
\setlength\belowdisplayskip{2pt}
\begin{align}
\mathcal{C}_{\rm{i}}=\left\{\left({\mathcal{R}}^{\rm{s}},{\mathcal{R}}^{\rm{c}}\right)|{\mathcal{R}}^{\rm{s}}\!\in\!\left[0,\mathcal{R}_{\rm{s}}^{\alpha}\right],
{\mathcal{R}}^{\rm{c}}\!\in\!\left[0,\mathcal{R}_{\rm{c}}^{\alpha}\right],\alpha\!\in\!\left[0,\!1\right]\right\}.\label{Rate_Regio_ISAC}
\end{align}
}\end{remark}
\vspace{-5pt}
\section{Performance of FDSAC}
We consider FDSAC as a baseline scenario, where $\kappa\in[0,1]$ fraction of the total bandwidth and $\mu\in[0,1]$ fraction of the total power is used for communications, and the other is used for sensing. Based on \cite{Ouyang2022_CL}, the CR and the SR are given by $\overline{\mathcal{R}}_{\rm{c}}^{\rm{f}}=\kappa\log_2(1+\frac{\mu}{\kappa} p\lVert{\mathbf{h}}_{\rm{c}}\rVert^2)$ and $\mathcal{R}_{\rm{s}}^{\rm{f}}=\frac{(1-\kappa)}{L}\log_2(1+\frac{1-\mu}{1-\kappa}pNL\alpha_{\rm{s}}\lVert{\mathbf{h}}_{\rm{s}}\rVert^2)$, respectively. Note that $(\overline{\mathcal{R}}_{\rm{c}}^{\rm{f}},\mathcal{R}_{\rm{s}}^{\rm{f}})$ can be analyzed in a similar way we analyze $({\overline{\mathcal{R}}}_{{\rm{c}}}^{\rm{c}},\mathcal{R}_{\rm{s}}^{\rm{s}})$.
\vspace{-5pt}
\begin{corollary}\label{FDSAC_Diversity_Order_FDSAC_HSPO}
As $p\rightarrow\infty$, a diversity order of $M$ is achievable for the OP $\Pr(\overline{\mathcal{R}}_{\rm{c}}^{\rm{f}}<\mathcal{R}_0)$ in the FDSAC system. Furthermore, the high-SNR slopes of $\mathcal{R}_{\rm{c}}^{\rm{f}}={\mathbbmss{E}}\{\overline{\mathcal{R}}_{\rm{c}}^{\rm{f}}\}$ and $\mathcal{R}_{\rm{s}}^{\rm{f}}$ are given by $\kappa$ and $(1-\kappa)\frac{1}{L}$, respectively.
\end{corollary}
\vspace{-5pt}
\vspace{-5pt}
\begin{corollary}\label{FDSAC_Rate_Region}
the rate region achieved by FDSAC is given by
{\setlength\abovedisplayskip{2pt}
\setlength\belowdisplayskip{2pt}
\begin{align}
\mathcal{C}_{\rm{f}}=\left\{\left({\mathcal{R}}^{\rm{s}},{\mathcal{R}}^{\rm{c}}\right)\left|
\begin{aligned}
&{\mathcal{R}}^{\rm{s}}\in\left[0,\mathcal{R}_{\rm{s}}^{\rm{f}}\right],
{\mathcal{R}}^{\rm{c}}\in\left[0,\mathcal{R}_{\rm{c}}^{\rm{f}}\right],\\
&\kappa\in\left[0,1\right],\mu\in\left[0,1\right]
\end{aligned}
\right.\right\}.\label{Rate_Regio_FDSAC}
\end{align}
}\end{corollary}
\vspace{-5pt}
After completing all the analyses, we summarize the results related to diversity order and high-SNR slope in Table \ref{table1}.
\vspace{-5pt}
\begin{remark}\label{High_SNR_Slope_Compare}
The results in Table \ref{table1} suggest that ISAC yields larger high-SNR slopes than FDSAC, which means that ISAC provides more degrees of freedom than FDSAC in terms of both communications and sensing.
\end{remark}
\vspace{-5pt}
We then compare the rate regions $\mathcal{C}_{\rm{i}}$ and $\mathcal{C}_{\rm{f}}$ as follows.
\vspace{-5pt}
\begin{theorem}\label{theorem_Rate_Region_Comparision}
The achievable rate regions satisfy $\mathcal{C}_{\rm{f}}\subseteq\mathcal{C}_{\rm{i}}$.
\end{theorem}
\vspace{-5pt}
\begin{IEEEproof}
Similar to the proof of \cite[Theorem 7]{Ouyang2023_WCL}.
\end{IEEEproof}
\vspace{-5pt}
\begin{remark}
The above results suggest that the rate region of FDSAC is entirely covered by that of ISAC.
\end{remark}
\vspace{-5pt}
\begin{table}[!t]
\centering
\setlength{\abovecaptionskip}{0pt}
\resizebox{0.26\textwidth}{!}{
\begin{tabular}{|c|cc|c|}
\hline
\multirow{2}{*}{System} & \multicolumn{2}{c|}{CR}                        & SR                       \\ \cline{2-4}
                        & \multicolumn{1}{c|}{$\mathcal{D}$} & $\mathcal{S}$ & $\mathcal{S}$            \\ \hline
C-C ISAC                    & \multicolumn{1}{c|}{$M$}    & $1$           & ${1}/{L}$           \\ \hline
Pareto Optimal ISAC         & \multicolumn{1}{c|}{$[1,M]$}    & $1$           & ${1}/{L}$           \\ \hline
S-C ISAC                    & \multicolumn{1}{c|}{$1$}    & $1$           & ${1}/{L}$           \\ \hline
FDSAC                   & \multicolumn{1}{c|}{$M$}    & $\kappa$    & $(1-\kappa)/{L}$ \\ \hline
\end{tabular}}
\caption{Diversity Order ($\mathcal{D}$) and High-SNR Slope ($\mathcal{S}$)}
\label{table1}
\vspace{-10pt}
\end{table}
\vspace{-10pt}
\section{Numerical Results}
In this section, the S\&C performance of ISAC is evaluated by using computer simulations. The parameters used for simulation are listed as follows: $M=4$, $N=5$, $L=20$, $\alpha_{\rm{c}}=1$, $\alpha_{\rm{s}}=1$, and ${\mathbf{b}}(\theta)=[{\rm{e}}^{{\rm{j}}\pi(m-1)\sin\theta}]_{m=1}^{M}$ with $\theta=0$.

\begin{figure}[!t]
    \centering
    \subfigbottomskip=0pt
	\subfigcapskip=-5pt
\setlength{\abovecaptionskip}{0pt}
    \subfigure[OP of the CR. $\kappa=\mu=0.5$.]
    {
        \includegraphics[height=0.16\textwidth]{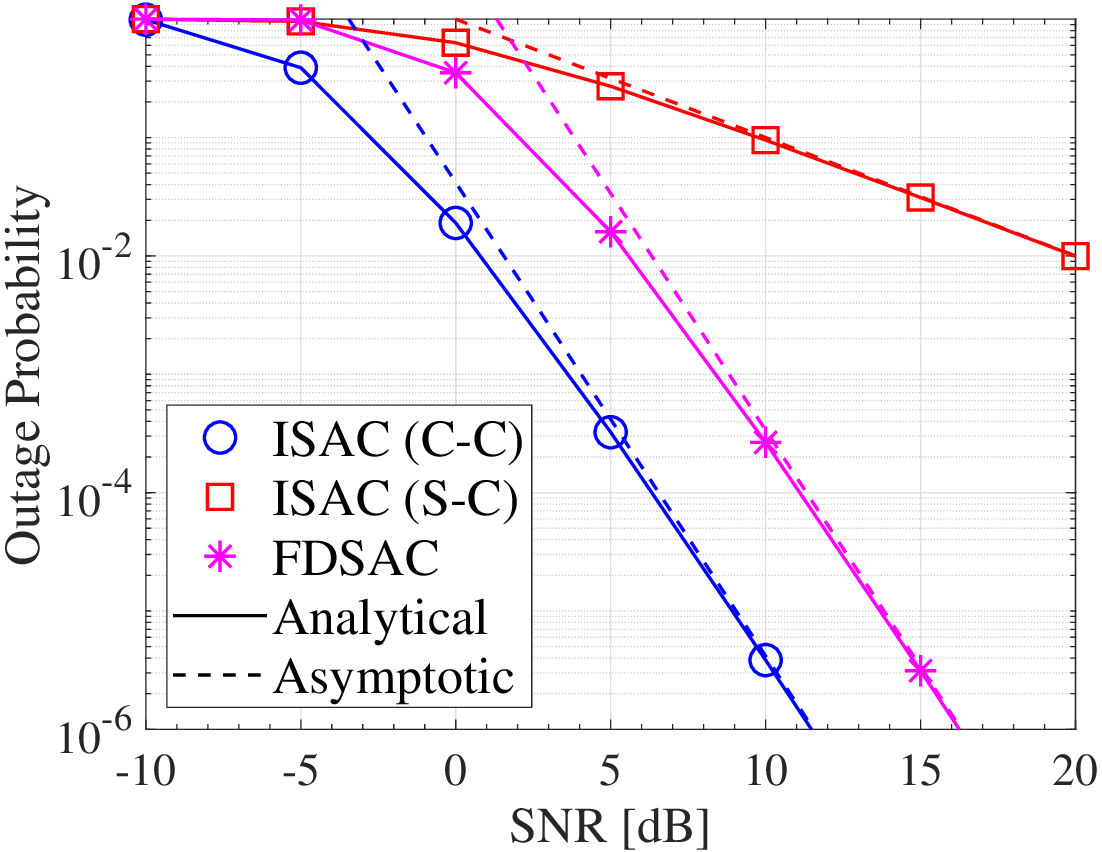}
	   \label{fig2a}	
    }
   \subfigure[ECR. $\kappa=\mu=0.5$.]
    {
        \includegraphics[height=0.16\textwidth]{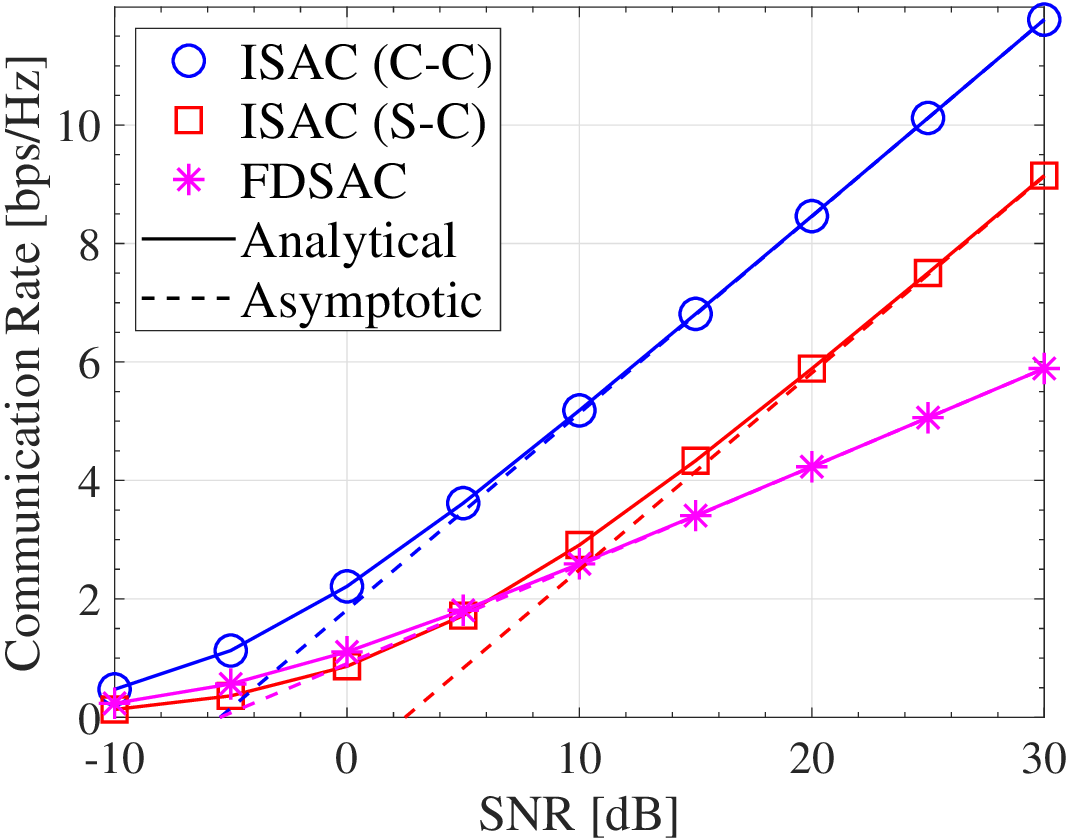}
	   \label{fig2b}	
    }
\caption{Performance of communications. ${\mathcal{R}}_0=1$ bps/Hz.}
    \label{Figure2}
    \vspace{-10pt}
\end{figure}

{\figurename} {\ref{fig2a}} depicts the OP as a function of the power budget $p$. The simulation results (symbols) align remarkably well with the analytical results, and in the high-SNR regime, the asymptotes accurately match the provided simulations. The graph clearly demonstrates that C-C ISAC achieves the lowest OP, followed by FDSAC and S-C ISAC. Additionally, it is observed that C-C ISAC and FDSAC have the same diversity order, which is higher than that of S-C ISAC. These findings are consistent with the conclusions drawn in Remark \ref{BF_Influence_Communications_1}. Turning to {\figurename} {\ref{fig2b}}, it showcases the ECR as a function of $p$. The analytical results exhibit a good match with the simulation data, and in the high-SNR region, the asymptotes precisely capture the behavior of the simulations. Notably, C-C ISAC achieves the largest ECR among the three cases considered. Moreover, S-C ISAC exhibits the same high-SNR slope as C-C ISAC, which is larger than that of FDSAC. Further observations reveal that when both C-C ISAC and S-C ISAC achieve the same ECR in the high-SNR region, C-C ISAC outperforms S-C ISAC by a constant power gap. In essence, this means that C-C ISAC yields a smaller high-SNR power offset compared to S-C ISAC. These results align with the findings presented in Remarks \ref{BF_Influence_Communications_1} and \ref{High_SNR_Slope_Compare}.

\begin{figure}[!t]
    \centering
    \subfigbottomskip=0pt
	\subfigcapskip=-5pt
\setlength{\abovecaptionskip}{0pt}
    \subfigure[$\kappa=\mu=0.5$.]
    {
        \includegraphics[height=0.16\textwidth]{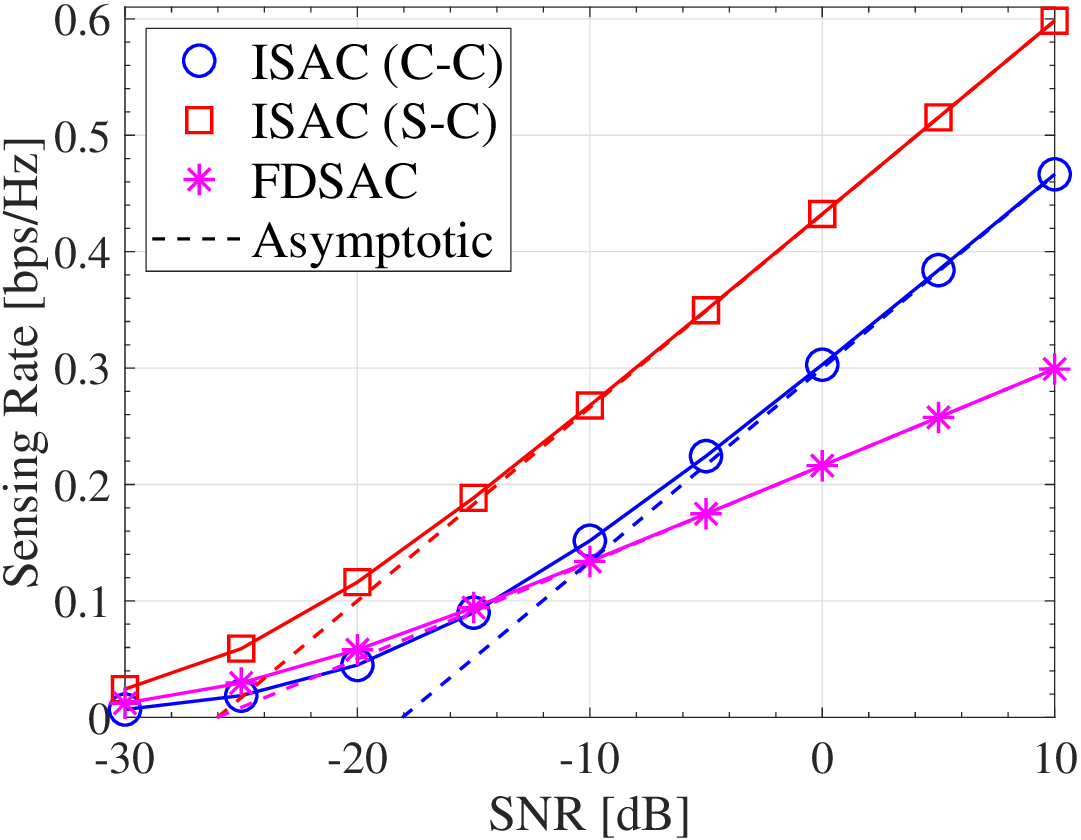}
	   \label{Figure3}	
    }
   \subfigure[$p=5$ dB.]
    {
        \includegraphics[height=0.16\textwidth]{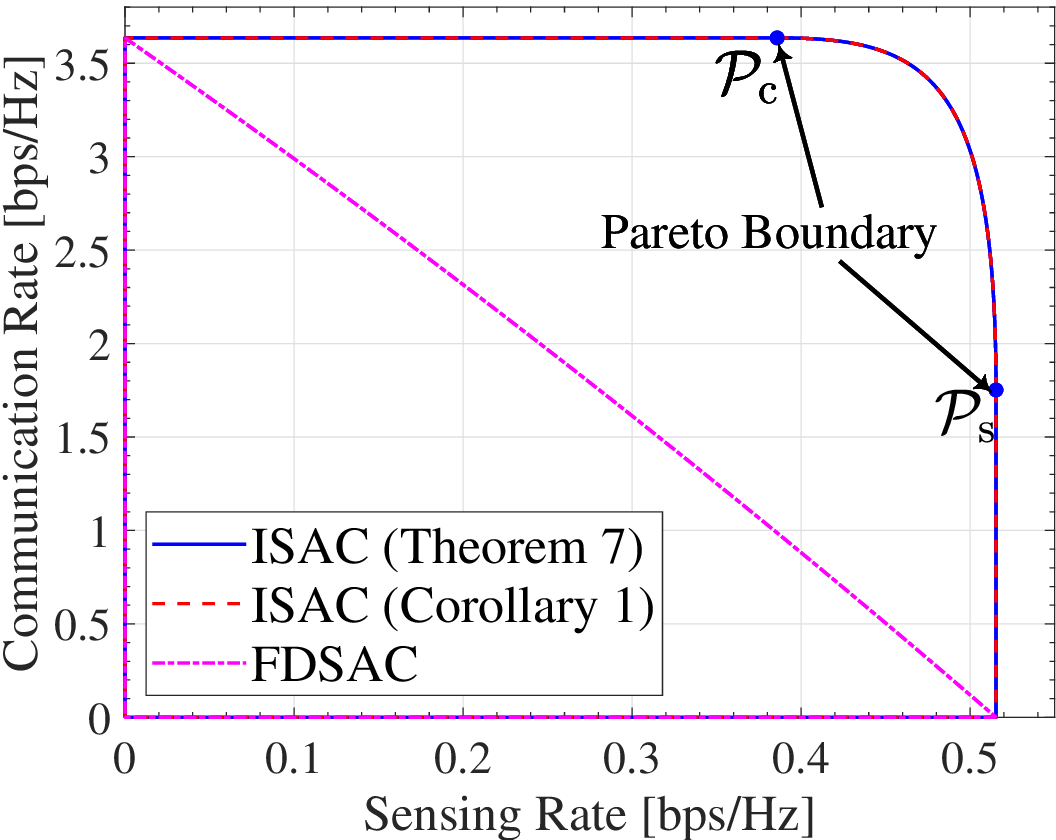}
	   \label{Figure4}	
    }
\caption{(a) Performance of sensing; (b) Rate region.}
    \vspace{-10pt}
\end{figure}

{\figurename} {\ref{Figure3}} displays the SR as a function of $p$. In the high-SNR region, the asymptotes closely match the provided simulations, ensuring the accuracy of our analysis. Notably, S-C ISAC achieves the highest SR, while FDSAC exhibits the lowest SR in the high-SNR regime. Furthermore, it is observed that S-C ISAC and C-C ISAC share the same high-SNR slope, with both being superior to FDSAC in this aspect, confirming the validity of Remark \ref{High_SNR_Slope_Compare}. Of particular interest is the fact that when achieving the same SR, S-C ISAC outperforms C-C ISAC by a constant power gap. This observation suggests that S-C ISAC yields a smaller high-SNR power offset compared to C-C ISAC, in line with the findings given in Remark \ref{BF_Influence_Sensing_1}.

In {\figurename} {\ref{Figure4}}, we present a comparison of the CR-SR region achieved by ISAC with that achieved by FDSAC. In the context of ISAC, two distinct points are of particular interest: point ${\mathcal{P}}_{\rm{s}}$, attained by the S-C design (i.e., ${\mathbf{w}}={\mathbf{w}}_0$ or ${\mathbf{w}}={\mathbf{w}}_{\rm{s}}$), and point ${\mathcal{P}}_{\rm{c}}$, achieved by the C-C design (i.e., ${\mathbf{w}}={\mathbf{w}}_1={\mathbf{w}}_{\rm{c}}$). Notably, the curve segment connecting ${\mathcal{P}}_{\rm{s}}$ and ${\mathcal{P}}_{\rm{c}}$ represents ISAC's Pareto boundary in terms of the rate region. Significantly, we make a noteworthy observation that FDSAC's rate region is entirely encompassed within ISAC's region. This finding unequivocally verifies the correctness of Theorem \ref{theorem_Rate_Region_Comparision}. Additionally, it is observed that the Pareto boundary achieved by the beamformer, as per Theorem \ref{Optimal_Solution_Problem_CR_SR_Tradeoff}, precisely aligns with that achieved by the beamformer, as per Corollary \ref{Simplified_Solution_Problem_CR_SR_Tradeoff}.
This alignment provides further support to the results presented in Remark \ref{Subspace_Trade_Off}.

\vspace{-10pt}
\section{Conclusion}
In this letter, we have proposed optimal beamforming design tailored to three distinct DFSAC scenarios. Novel expressions have been derived to characterize the achieved OP, ECR, and SR. Theoretical analyses and numerical simulations have shown that the DFSAC beamforming design influences SR and CR by shaping the high-SNR power offsets and diversity orders. Besides, it has been demonstrated that ISAC achieves a broader rate region than FDSAC.
\vspace{-10pt}
\begin{appendix}
\subsection{Proof of Lemma \ref{SR_Def_Lemma}}\label{Proof_SR_Def_Lemma}
Inserting \eqref{Target_Response_Matrix_Formula} into \eqref{reflected_echo_signal_matrix} gives ${\mathbf{Y}}_{\rm{s}}=
\sqrt{p}{\mathbf{b}}^{\mathsf{H}}\left(\theta\right){\mathbf{w}}{\mathbf{a}}\left(\theta\right)\mathbf{s}^{\mathsf{H}}\beta+{\mathbf{N}}_{\rm{s}}$, which satisfies ${\mathsf{vec}}({\mathbf{Y}}_{\rm{s}})=\sqrt{p}{\mathbf{b}}^{\mathsf{H}}\left(\theta\right){\mathbf{w}}{\mathsf{vec}}({\mathbf{a}}\left(\theta\right)\mathbf{s}^{\mathsf{H}})\beta+
{\mathsf{vec}}({\mathbf{N}}_{\rm{s}})$.Therefore, the conditioned MI between ${\mathsf{vec}}({\mathbf{Y}}_{\rm{s}})$ (or ${\mathbf{Y}}_{\rm{s}}$) and $\beta$ can be treated as the capacity of a multiple-input single-input Gaussian channel with Gaussian distributed inputs $\beta\sim{\mathcal{CN}}(0,\alpha_{\rm{s}})$ and channel vector $\sqrt{p}{\mathbf{b}}^{\mathsf{H}}\left(\theta\right){\mathbf{w}}{\mathsf{vec}}({\mathbf{a}}\left(\theta\right)\mathbf{s}^{\mathsf{H}})$. Thus, the sensing MI satisfies $I\left({\mathbf{Y}}_{\rm{s}};\beta|\mathbf{X}\right)=\log_2(1+p\lVert{\mathbf{s}}\rVert^2\lVert{\mathbf{a}}(\theta)\rVert^2\alpha_{\rm{s}}\lvert{\mathbf{w}}^{\mathsf{H}}{\mathbf{h}}_{\rm{s}}\rvert^2)$. The final results can be obtained by using the fact that $\lVert{\mathbf{s}}\rVert^2=L$ and $\lVert{\mathbf{a}}(\theta)\rVert^2=N$.
\vspace{-10pt}
\subsection{Proof of Theorem \ref{Ergodic_Communication_Rate_C_C_Theorem}}\label{Proof_Ergodic_Communication_Rate_C_C_Theorem}
Under the C-C design, we have $\gamma_{\rm{c}}=p\lVert{\mathbf{h}}_{\rm{c}}\rVert^2$. Since ${\mathbf{h}}_{\rm{c}}\sim{\mathcal{CN}}({\mathbf{0}},\alpha_{\rm{c}}{\mathbf{I}})$, the PDF and the CDF of $\lVert{\mathbf{h}}_{\rm{c}}\rVert^2$ are given by $f_{\rm{c}}(x)=\frac{x^{M-1}}{\Gamma(M)\alpha_{\rm{c}}^{M}}{\rm{e}}^{-\frac{x}{\alpha_{\rm{c}}}}$ and $F_{\rm{c}}(x)=\frac{1}{\Gamma(M)}\gamma(M,\frac{x}{\alpha_{\rm{c}}})$, respectively. Under the S-C design, we have $\gamma_{\rm{c}}=p\lvert{\mathbf{w}}_{\rm{s}}^{\mathsf{H}}{\mathbf{h}}_{\rm{c}}\rvert^2$. Since ${\mathbf{w}}_{\rm{s}}$ is independent with ${\mathbf{h}}_{\rm{c}}$, we have ${\mathbf{w}}_{\rm{s}}^{\mathsf{H}}{\mathbf{h}}_{\rm{c}}\sim{\mathcal{CN}}(0,\alpha_{\rm{c}})$. The resulting PDF and CDF of $\lvert{\mathbf{w}}_{\rm{s}}^{\mathsf{H}}{\mathbf{h}}_{\rm{c}}\rvert^2$ can be obtained by replacing $M$ in $f_{\rm{c}}(\cdot)$ and $F_{\rm{c}}(\cdot)$ with $1$, respectively. After obtaining the PDF and CDF, we can derive the closed-form expressions of the OP, ECR, and their high-SNR approximations by using similar steps as those outlined in \cite[Appendices A--B]{Ouyang2022_CL}.
\vspace{-10pt}
\subsection{Proof of Theorem \ref{Average_Sensing_Rate_C_C_Theorem}}\label{Proof_Average_Sensing_Rate_C_C_Theorem}
Using the fact that $\lim_{x\rightarrow\infty}\log_2(1+x)\approx\log_2{x}$, we obtain $\lim_{p\rightarrow\infty}{{\mathcal R}}_{\rm{s}}^{\rm{c}}\approx L^{-1}\log_2(pNL\alpha_{\rm{s}})+L^{-1}{\mathbbmss{E}}
\{\log_2(\lvert{\mathbf{h}}_{\rm{c}}^{\mathsf{H}}{\mathbf{h}}_{\rm{s}}\rvert^2)\}-L^{-1}{\mathbbmss{E}}\{\log_2({\lVert{\mathbf{h}}_{\rm{c}}\rVert}^{2})\}$. It follows from ${\mathbf{h}}_{\rm{c}}\sim{\mathcal{CN}}({\mathbf{0}},\alpha_{\rm{c}}{\mathbf{I}})$ that ${\mathbf{h}}_{\rm{c}}^{\mathsf{H}}{\mathbf{h}}_{\rm{s}}\sim{\mathcal{CN}}(0,\alpha_{\rm{c}}{\lVert{\mathbf{h}}_{\rm{s}}\rVert}^{2})$. The final results can thus be derived by using \cite[Eq. (4.352.1)]{Ryzhik2007} and the fact that $\psi(x)=\psi(1)+\sum_{i=1}^{x-1}\frac{1}{i}$ for $x\in{\mathbbmss{Z}}^{+}$.
\vspace{-10pt}
\subsection{Proof of Theorem \ref{Optimal_Solution_Problem_CR_SR_Tradeoff}}\label{Proof_Optimal_Solution_Problem_CR_SR_Tradeoff}
The optimal solution to problem \eqref{Problem_CR_SR_Tradeoff} can be obtained from the Karush-Kuhn-Tucker (KKT) condition as follows:
{\setlength\abovedisplayskip{2pt}
\setlength\belowdisplayskip{2pt}
\begin{numcases}{}
  \nabla(-{\mathcal{R}})+\lambda\nabla(\lVert{\mathbf{w}}\rVert^2\!-\!1)+\mu_1\nabla f_{1}+\mu_2\nabla f_{2}={\mathbf{0}}, \label{KKT_1} \\
  \mu_1f_1=0, \mu_2f_2=0, \lambda\in{\mathbbmss{R}},\mu_1\geq0,\mu_2\geq0, \label{KKT_2}
\end{numcases}
}where $f_{i}={\rm{e}}^{\beta_i{\mathcal{R}}}-1-{\mathbf{w}}^{\mathsf{H}}{\mathbf{h}}_i{\mathbf{h}}_i^{\mathsf{H}}{\mathbf{w}}$ for $i\in\{1,2\}$, and $\{\lambda,\mu_1,\mu_2\}$ are real-valued Lagrangian multipliers. From \eqref{KKT_1}, it can be shown that
{\setlength\abovedisplayskip{2pt}
\setlength\belowdisplayskip{2pt}
\begin{numcases}{}
(\mu_1{\mathbf{h}}_1{\mathbf{h}}_1^{\mathsf{H}}+\mu_2{\mathbf{h}}_2{\mathbf{h}}_2^{\mathsf{H}}){\mathbf{w}}=\lambda{\mathbf{w}},\label{KKT_1_Dev1}\\
\mu_1{\rm{e}}^{\beta_1{\mathcal{R}}}\beta_1+\mu_2{\rm{e}}^{\beta_2{\mathcal{R}}}\beta_2=1\label{KKT_1_Dev2}.
\end{numcases}
}It is clear that the optimal beamformer ${\mathbf{w}}$ is an eigenvector of the matrix $(\sum_{i=1}^{2}\mu_i{\mathbf{h}}_i{\mathbf{h}}_i^{\mathsf{H}})$ with a corresponding eigenvalue $\lambda$. It follows from \eqref{KKT_1_Dev2} that $\mu_1$ and $\mu_2$ cannot be $0$ at the same time. Moreover, from \eqref{KKT_1_Dev1}, we have
{\setlength\abovedisplayskip{2pt}
\setlength\belowdisplayskip{2pt}
\begin{equation}\label{KKT_2_Dev1}
\mu_1{\mathbf{w}}^{\mathsf{H}}{\mathbf{h}}_1{\mathbf{h}}_1^{\mathsf{H}}{\mathbf{w}}+
\mu_2{\mathbf{w}}^{\mathsf{H}}{\mathbf{h}}_2{\mathbf{h}}_2^{\mathsf{H}}{\mathbf{w}}=\lambda{\mathbf{w}}^{\mathsf{H}}{\mathbf{w}}=\lambda.
\end{equation}
}The above results suggest that $\lambda>0$. It is widely known that the eigenvector of $(\sum_{i=1}^{2}\mu_i{\mathbf{h}}_i{\mathbf{h}}_i^{\mathsf{H}})$ corresponding a non-zero eigenvalue can be written as the linear combination of ${\mathbf{h}}_1$ and $\mathbf{h}_2$. Thus, ${\mathbf{w}}=a{\mathbf{h}}_1+b{\mathbf{h}}_2$. Particularly, we have ${\mathbf{w}}={\mathbf{w}}_{\rm{c}}$ for $\mu_2=0$ and $\mu_1>0$, and ${\mathbf{w}}={\mathbf{w}}_{\rm{s}}$ for $\mu_1=0$ and $\mu_2>0$. Consider the case of $\mu_1>0$ and $\mu_2>0$, which yields
{\setlength\abovedisplayskip{2pt}
\setlength\belowdisplayskip{2pt}
\begin{equation}\label{KKT_1_Dev3}
f_i=0\Leftrightarrow{\rm{e}}^{\beta_i{\mathcal{R}}}-1={\mathbf{w}}^{\mathsf{H}}{\mathbf{h}}_i{\mathbf{h}}_i^{\mathsf{H}}{\mathbf{w}},~i=1,2.
\end{equation}
}Substituting ${\mathbf{w}}=a{\mathbf{h}}_1+b{\mathbf{h}}_2$ into \eqref{KKT_1_Dev1} and \eqref{KKT_1_Dev3} and performing some simple mathmematical manipulations, we have
{\setlength\abovedisplayskip{2pt}
\setlength\belowdisplayskip{2pt}
\begin{numcases}{}
a/b={(\mu_1\sqrt{{{\rm{e}}^{\beta_1{\mathcal{R}}}\!\!-\!\!1}})}/
{(\mu_2\sqrt{{{\rm{e}}^{\beta_2{\mathcal{R}}}\!\!-\!\!1}}{\rm{e}}^{-{\rm{j}}\angle\alpha_{12}})},\label{KKT_1_Dev4}\\
\mu_1(\alpha_{11}+\alpha_{12}b/a)=\mu_2(\alpha_{12}^{*}a/b+\alpha_{22})=\lambda.\label{KKT_1_Dev5}
\end{numcases}
}By combining \eqref{KKT_1_Dev2} and \eqref{KKT_1_Dev5}, we have $\mu_1=\frac{\varrho_1}{\chi}$ and $\mu_2=\frac{\varrho_2}{\chi}$. Inserting $\mu_1=\frac{\varrho_1}{\chi}$ and $\mu_2=\frac{\varrho_2}{\chi}$ into \eqref{KKT_1_Dev5} gives $\lambda=(\alpha_{11}\alpha_{22}-|\alpha_{12}|^2)/\chi$. Furthermore, substituting the above results and \eqref{KKT_1_Dev3} into \eqref{KKT_2_Dev1} gives $\alpha_{11}\alpha_{22}-|\alpha_{12}|^2=(\alpha_{11}-\delta^{-1}|\alpha_{12}|)({\rm{e}}^{\beta_2{\mathcal{R}}}-1)
+(\alpha_{22}-\delta|\alpha_{12}|)({\rm{e}}^{\beta_1{\mathcal{R}}}-1)$. When $\mu_1>0$ and $\mu_2>0$, it is easily shown that $(\alpha_{11}-\delta^{-1}|\alpha_{12}|)({\rm{e}}^{\beta_2{\mathcal{R}}}-1)
+(\alpha_{22}-\delta|\alpha_{12}|)({\rm{e}}^{\beta_1{\mathcal{R}}}-1)$ is a monotonic function with $\mathcal{R}$. By solving this equation, we can obtain the optimal solution of $\mathcal{R}$, and the final results follow immediately.
\vspace{-10pt}
\subsection{Proof of Corollary \ref{Simplified_Solution_Problem_CR_SR_Tradeoff}}\label{Proof_Simplified_Solution_Problem_CR_SR_Tradeoff}
Let $({\mathcal{R}}_{\alpha}^{\rm{s}},{\mathcal{R}}_{\alpha}^{\rm{c}})$ and $({\mathcal{R}}_{\xi}^{\rm{s}},{\mathcal{R}}_{\xi}^{\rm{c}})$ denote the SR-CR pairs achieved by ${\mathbf{w}}_\alpha^{\star}$ and ${\mathbf{w}}_{\xi}$, respectively. It is easily shown that ${\mathcal{R}}_{\xi}^{\rm{s}}$ decreases with $\xi$, whereas ${\mathcal{R}}_{\xi}^{\rm{c}}$ increases with $\xi$. Let $({\mathcal{R}}_{1},{\mathcal{R}}_{2})$ denote the achievable SR-CR pair. Thus, the attainable rate regions achieved by ${\mathbf{w}}_\alpha^{\star}$ and ${\mathbf{w}}_{\xi}$ are given by
{\setlength\abovedisplayskip{2pt}
\setlength\belowdisplayskip{2pt}
\begin{align}
&\mathcal{C}_{1}=\left\{\left({\mathcal{R}}_{1},{\mathcal{R}}_{2}\right)|{\mathcal{R}}_{\rm{s}}\!\in\!\left[0,\mathcal{R}_{\alpha}^{\rm{s}}\right],\!
{\mathcal{R}}_{\rm{c}}\!\in\!\left[0,\mathcal{R}_{\alpha}^{\rm{c}}\right],\alpha\!\in\!\left[0,\!1\right]\right\},\\
&\mathcal{C}_{2}=\left\{\left({\mathcal{R}}_{1},{\mathcal{R}}_{2}\right)|{\mathcal{R}}_{\rm{s}}\!\in\!\left[0,\mathcal{R}_{\xi}^{\rm{s}}\right]\!,\!
{\mathcal{R}}_{\rm{c}}\!\in\!\left[0,\mathcal{R}_{\xi}^{\rm{c}}\right]\!,\xi\!\in\!\left[0,\!1\right]\right\},
\end{align}
}respectively. Since $\mathcal{C}_{1}$ contains all the achievable rate pairs, we have ${\mathcal{C}}_2\subseteq{\mathcal{C}}_1$. Besides, since ${\mathbf{w}}_\alpha^{\star}$ is the linear combination of ${\mathbf{h}}_{\rm{c}}$ and ${\mathbf{h}}_{\rm{s}}{\rm{e}}^{-{\rm{j}}\angle\alpha_{12}}$ with non-negative real coefficients, we have ${\mathcal{C}}_1\subseteq{\mathcal{C}}_2$. It follows that ${\mathcal{C}}_2={\mathcal{C}}_1$, which means that ${\mathcal{C}}_2$ and ${\mathcal{C}}_1$ have the same boundary. The final results follow directly.
\vspace{-10pt}
\end{appendix}

\end{document}